\documentclass[PRO,english]{ipsj}
\usepackage{PROpresentation}
\PROheadtitle{}

\usepackage{balance}  
\usepackage{color}    
\usepackage{url} 
\usepackage{amssymb}
\usepackage{amsmath}
\usepackage{caption}
\usepackage{listings}
\usepackage{paralist} 
\usepackage{llvm/lang} 
\usepackage{nasm/style} 
\lstdefinestyle{mlir-code}{
  basicstyle=\fontsize{12}{12}\ttfamily,
  language=llvm, style=nasm,
  numbers=left, stepnumber=1, xleftmargin=2.5em,
  float=t,
  breaklines=true, breakatwhitespace=true,
  escapechar=\& 
}

\lstdefinestyle{mlir-code-inpara}{
  basicstyle=\fontsize{12}{12}\ttfamily,
  language=llvm, style=nasm,
  numbers=left, stepnumber=1, xleftmargin=1em,
  aboveskip=5pt, belowskip=5pt,
  breaklines=true, breakatwhitespace=true,
  escapechar=\&, 
  frame=none, 
}

\newcommand{\dialect}[1]{{\sf #1}}
\newcommand{\op}[1]{{\sf #1}}
\newcommand{\shape}[1]{${\sf \langle#1\rangle}$}
\newcommand{\tensor}[1]{${\sf tensor\langle#1\rangle}$}
\newcommand{\memref}[1]{${\sf memref\langle#1\rangle}$}
\newcommand{\diop}[2]{{\sf #1}.{\sf #2}}  
\newcommand{\pass}[1]{{\sf -{}-{#1}}} 
\newcommand{\onnxmlir}{{\it onnx-mlir}}

\usepackage{booktabs}   

\usepackage{graphicx}
\graphicspath{ {./images/} }

\begin{document}

\title{Compiling ONNX Neural Network Models Using MLIR}

\affiliate{WATSON}{IBM T.J. Watson Research Center,\\
1101 Kitchawan Rd, Yorktown Heights, NY 10598, USA}

\affiliate{TRL}{IBM Research - Tokyo,\\
19-21, Nihonbashi Hakozaki-cho, Chuo-ku, Tokyo 103-8510, Japan}

\paffiliate{PMIT}{Massachusetts Institute of Technology (MIT)}

\cauthor{CAUTHOR}{Tung D. Le (tung@jp.ibm.com), Tian Jin (tianjin@mit.edu)}

\author{Tian Jin}{WATSON, PMIT, CAUTHOR}[]
\author{Gheorghe-Teodor Bercea}{WATSON}[]
\author{Tung D. Le}{TRL, CAUTHOR}[]
\author{Tong Chen}{WATSON}[]
\author{Gong Su}{WATSON}[]
\author{Haruki Imai}{TRL}[]
\author{Yasushi Negishi}{TRL}[]
\author{Anh Leu}{WATSON}[]
\author{Kevin O'Brien}{WATSON}[]
\author{Kiyokuni Kawachiya}{TRL}[]
\author{Alexandre E. Eichenberger}{WATSON}[]

\begin{abstract}
  Deep neural network models are becoming increasingly popular and have been used in various tasks such as computer vision, speech recognition, and natural language processing.
Machine learning models are commonly trained in a resource-rich environment and then deployed in a distinct environment such as high availability machines or edge devices.
To assist the portability of models, the open-source community has proposed the Open Neural Network Exchange (ONNX) standard.
In this paper, we present a high-level, preliminary report on our onnx-mlir compiler, which generates code for the inference of deep neural network models described in the ONNX format.
Onnx-mlir is an open-source compiler implemented using the Multi-Level Intermediate Representation (MLIR) infrastructure recently integrated in the LLVM project.
Onnx-mlir relies on the MLIR concept of dialects to implement its functionality.
We propose here two new dialects: 
\begin{inparaenum}
\item an ONNX specific dialect that encodes the ONNX standard semantics, and
\item a loop-based dialect to provide for a common lowering point for all ONNX dialect operations.
\end{inparaenum}
Each intermediate representation facilitates its own characteristic set of graph-level and loop-based optimizations respectively.
We illustrate our approach by following several models through the proposed representations and we include some early optimization work and performance results.

\end{abstract}

\maketitle

\section{Introduction}
\label{sec:intro}
Deep neural network models have been used widely for various tasks such as computer vision, speech recognition, and natural language processing.
The success of such models was mainly originated from the development of accelerators, especially GPU accelerators, back in 2012~\cite{nips2012:alexnet}.
Since then, many deep learning frameworks, such as Torch, Caffe, Theano, and TensorFlow, have been developed to facilitate the training and inferencing of deep neural network models, which significantly speeds up the explosion of deep learning in many areas.
However, training and inferencing are often done on different environments due to their different optimization characteristics.
For example, a model is trained using a large-scale distributed system since it might need weeks or months to finish, and can then be used on light-weight devices such as Internet of Things or mobile phones for inferencing. 
Hence, it is desirable to dynamically rewrite a trained model so that it runs efficiently on a target environment.

Many deep learning frameworks utilize a highly-optimized library written for a target accelerator.
Rewriting a model for inferencing consists of replacing the operations in the model with the function calls in the library.
While such a library-call approach simplifies the rewritten procedure and would lead to improved performance, it exposes the following drawbacks. 
Firstly, the number of models that can be rewritten is limited by the provided functions in the library. 
Secondly, it is often the case that users need to install additional packages to make the library work well.
Thirdly, it lacks the ability to tailor code specific to different problems since the same function may be used for them.

We tackle these drawbacks by developing a compiler that rewrites a trained model to native code for a target hardware.
It uses many mature optimization techniques developed during the long history of compiler, such as the ability to tailor code for a specific problem, memory optimizations, and parallelization.
Our compiler is completely based on open-source software.
In particular, we chose Open Neural Network Exchange (ONNX)~\cite{bai2020:onnx} as a format to represent the input model of our compiler.
ONNX is an open-source machine-independent format and widely used for exchanging neural network models.
It has been actively maintained by and contributed from open source communities.
Our compiler was written using Multi-level Intermediate Representation (MLIR)~\cite{lattner2020:mlir}, a modern open source compiler infrastructure for multi-level intermediate representations and a subproject inside LLVM~\cite{cgo04:llvm}.

Our compiler is completely open-sourced and a subproject inside the ONNX project\footnote{\url{https://github.com/onnx/onnx-mlir}}.
Although it is still under development, it can already compile some popular models such MNIST and ResNet50 to native code on x86 machines, IBM Power Systems\footnote{\url{https://www.ibm.com/it-infrastructure/power/power9}}, and IBM System Z\footnote{\url{https://www.ibm.com/it-infrastructure/z/hardware}}.
In this paper, we will introduce our compiler by
\begin{itemize}
  \item presenting its overall design and architecture of the compiler,
  \item introducing two new dialects: \dialect{onnx} dialect to encode the ONNX standard semantics, and \dialect{krnl} dialect to provide for a common lowering point for all ONNX dialect operations.
  \item introducing optimization passes such as graph rewriting, constant propagation, and memory management, and
  \item discussing some problems we encountered when emitting native code for different architectures.
\end{itemize}

The remainder of the paper is organized as follows.
In Sec.~\ref{sec:mlir}, we briefly discuss ONNX and MLIR on which our compiler is based.
In Sec.~\ref{sec:onnx-mlir}, we introduce our compiler, its design principle, and architecture. 
We also discuss in this section two new dialects, i.e., \dialect{onnx} and \dialect{krnl}, and some optimization passes.
In Sec.~\ref{sec:experiment}, we present some preliminary experiemental results for MNIST and ResNet50 models on IBM Power Systems.
Finally, we conclude our paper and discuss future work in Sec.~\ref{sec:conclusion}.

\section{Background}
\label{sec:mlir}
\begin{lstlisting}[style=mlir-code, caption={ONNX model for LeakyRelu operator (printed using `protoc' command).}, label={lst:proto-exp}]
ir_version: 3
producer_name: "backend-test"
graph {
  node {
    input: "x"
    output: "y"
    op_type: "LeakyRelu"
    attribute {
      name: "alpha"
      f: 0.1
      type: FLOAT
    }
  }
  name: "test_leakyrelu"
  input {
    name: "x"
    type {
      tensor_type {
        elem_type: 1
        shape {
          dim {
            dim_value: 3
          }
          dim {
            dim_value: 4
          }
          dim {
            dim_value: 5
          }
        }
      }
    }
  }
  output {
    name: "y"
    type {
      tensor_type {
        elem_type: 1
        shape {
          dim {
            dim_value: 3
          }
          dim {
            dim_value: 4
          }
          dim {
            dim_value: 5
          }
        }
      }
    }
  }
}
opset_import {
  version: 9
}
\end{lstlisting}

\subsection{ONNX}
Open Neural Network Exchange (ONNX)~\cite{bai2020:onnx} is an open source format for artificial intelligence models, including both deep learning and traditional machine learning.
It defines an extensible computational graph model, operators, and standard data types, which provides a common IR for different frameworks.
There are two ONNX variants: the neural-network-only ONNX variant recognizes only tensors as input and output types, while the classic machine learning ONNX-ML also recognizes sequences and maps.  
ONNX-ML extends the ONNX operator set with machine learning algorithms that are not based on neural networks.
In this paper, we focus on the neural-network-only ONNX variant and refer to it as just ONNX.

In ONNX, the top-level structure is a `Model' to associate metadata with a graph. 
Operators in ONNX are divided into a set of primitive operators and functions, where a function is an operator whose calculation can be expressed via a subgraph of other operators.
A graph is used to describe a function.
There are lists of nodes, inputs, outputs, and initializers (constant values or default values for inputs) in a graph. 
An acyclic dataflow graph is constructed as a topological sort of the list of nodes in the graph.
Each node in a graph contains the name of the operator it invokes, inputs, outputs, and attributes associated with the operator.
Inputs and outputs can be marked as variadic or optional.
There are three data types used to define inputs and outputs, i.e., `Tensor', `Sequence', and `Map'.

ONNX uses the Protocol Buffers\footnote{https://developers.google.com/protocol-buffers} definition language for its syntax.
Listing~\ref{lst:proto-exp} shows an example of an ONNX model for the LeakyRelu operator.
There is one node in the graph (Lines~$4$--$13$), which is associated with LeakyRelu, and has one input, one output, and one attribute.
The input and output tensors have the shape of \shape{3x4x5} and element type of float32 ({\sf elem\_type: 1} at Lines~$19$ and $38$). 

\begin{figure}[t]
   \centering
   \includegraphics[scale=0.54]{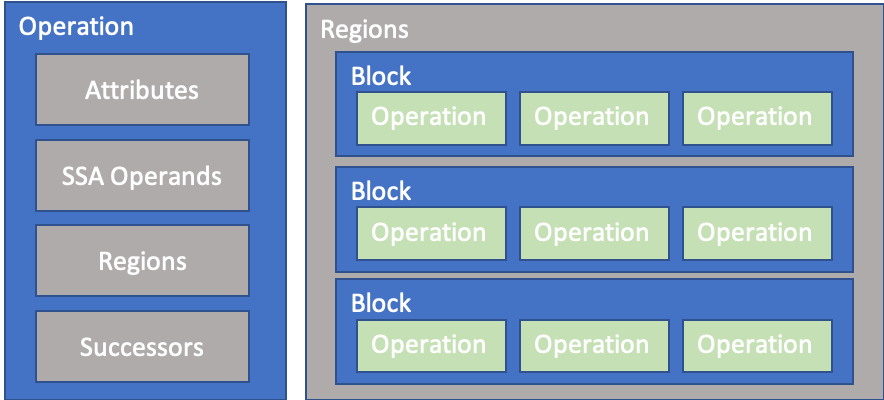}
   \caption{Operations and Regions in MLIR.}
   \label{fig:mlir-operation}
\end{figure}

\begin{lstlisting}[style=mlir-code, caption={Compute the exponential of a tensor in MLIR.}, label={lst:mlir-exp}, float=*t]
module {
  func @exp(arg0: memref<3x4xf32>) -> memref<3x4xf32> {
    %1 = std.alloc() : memref<3x4xf32>
    affine.for %arg1 = 0 to 3 {
      affine.for %arg2 = 0 to 4 {
        %2 = affine.load %arg0[%arg1, %arg2] : memref<3x4xf32>
        %3 = std.exp %2 : f32
        affine.store %3, %1[%arg1, %arg2] : memref<3x4xf32>
      }
    }
    std.return %1 : memref<3x4xf32>
  }
}
\end{lstlisting}

\subsection{MLIR}
Multi-level Intermediate Representation (MLIR)~\cite{lattner2020:mlir} is a modern compiler infrastructure which is reusable and extensible.
It reduces the cost of building domain-specfic compilers by facilitating the design and implementation of code generators, translators, and optimizers at different abstraction levels.
MLIR is a subproject of the LLVM project~\cite{llvm-project} and has many similarities to the LLVM compiler infrastructure~\cite{cgo04:llvm}.
In this section, we briefly review some of the features in MLIR that were used to build our compiler.
For more information about MLIR, one can refer to a previous study~\cite{lattner2020:mlir}.
Readers who are familiar with MLIR can skip this section.

Similar to LLVM, MLIR is a three-address static single assignment (SSA)-based IR, where values are defined before use and have a scope defined by their dominance relations.
Operations may produce zero or more results, and each operation is a distinct SSA value with its own type defined by the type system.
The type system in MLIR is open, and one can define application-specific types.
There are a number of primitive types, e.g., integers, as well as aggregate types for tensors and memory buffers, e.g., `Tensor' and `MemRef' types.
A Tensor type is abstracted and does not have a pointer to the data while a MemRef type is a lower representation, referring to a region of memory.
In MLIR, Tensor and MemRef types are syntactically represented as \tensor{D_1{\times}D_2{\times}\ldots{\times}D_N{\times}\sf{dtype}} and \memref{D_1{\times}D_2{\times}\ldots{\times}D_N{\times}\sf{dtype}}, respectively,
where $D_1, D_2, \ldots, D_N$ are intergers representing the dimensions of a tensor or memref,
and {\sf dtype} is the type of the elements in a tensor or memref, e.g., {\sf f32} for float32.
\shape{D_1{\times}D_2{\times}\ldots{\times}D_N} is called the shape of a tensor or memref. 
Tensor and MemRef types can be unranked when their shapes are unknown.
In MLIR, unranked Tensor and MemRef types are syntactically represented as \tensor{*{\times}\sf{dtype}} and \memref{*{\times}\sf{dtype}}, respectively.

An \emph{operation} is the unit of code in MLIR.
To define an operation, a TableGen-based~\cite{llvm-tablegen} specification for an operation descriptor is used.
Figure~\ref{fig:mlir-operation} shows the structure of an operation.
An operation has a list of SSA operands and may have attributes that store static information.
An operation can hold a \emph{region} which is a list of \emph{blocks}.
A \emph{block} contains a list of operations and ends with a \emph{terminator} operation that may have \emph{successor} blocks to which the control flow may be transferred.
That being said, \emph{nested regions} becomes a first-class concept in MLIR, which is efficient to represent control flow graphs.
A \emph{function} is an operation with a single region and attributes.
A \emph{module} is an operation with a single region containing a single block and terminated by a dummy operation.

To develop a compiler using MLIR, users often need to define \emph{dialects} and \emph{optimization passes}.
A dialect serves as an abstraction level or intermediate representation,
and an optimization pass is to enable optimization at an abstraction level or transformation among abstraction levels.

There are dialects in MLIR that are ready to use, e.g., \dialect{llvm}, \dialect{std}, \dialect{scf}, and \dialect{affine}.
The \dialect{llvm} dialect is a low-level dialect.
It wraps the LLVM IR types and instructions into MLIR types and operations.
The \dialect{std} dialect includes standard operations such as \op{load}, \op{store}, \op{addi}, \op{addf}, \op{absf}, and \op{call}.
The \dialect{scf} dialect defines control flow operations such as \op{for} and \op{if}.
The \dialect{affine} dialect provides an abstraction for affine operations and analyses.

Optimization passes can be roughly classified into three categories: general transformation, conversion, and dialect-specific.
General transformation passes includes common passes such as `canonicalize' pass for operation canonicalization, `CSE' pass to eliminate common sub-expressions, and passes to print IR information such as `print-op-graph', `print-op-stats', and `print-cfg-graph'.
Conversion passes are to convert operations in one dialect to operations in another dialect, e.g., `convert-std-to-llvm' pass to convert standard operations into LLVM instructions.
Finally, dialect-specific passess are for transformation in a dialect, e.g., `affine-loop-unroll-jam' pass to unroll and jam affine loops in the \dialect{affine} dialect.
MLIR passes can be expressed via Declarative Rewriting Rules (DRRs) using tablegen records or via writing code in C++.

To denote an operation in a dialect, we explicitly use a form of \diop{dialect\_name}{operation\_name}.
For example, \diop{std}{load} means the operation \op{load} of dialect \dialect{std}.
Optimization passes are named with prefix `\pass{}', for example, \pass{canonicalize} is the canonlicalization pass.

Listing~\ref{lst:mlir-exp} shows an example for calculating the exponential of a given input tensor, element-wise, using \dialect{std} and \dialect{affine} dialects.
The top level is a module containing a function `exp'.
The function `exp' accepts one input that is of \op{memref} type, and produces an output of the same type.
The memory for the output is allocated via \diop{std}{alloc} (Line~$3$).
There is a nested loop (Lines~$4$--$10$), iterating over dimensions of the inputs using \diop{affine}{for}, loading each element from the input using \diop{affine}{load} (Line~$6$), computing the exponential using \diop{std}{exp} (Line~$7$), and storing the result in the output using \diop{affine}{store} (Line~$8$).
The output of the function is finally returned using \diop{std}{return}.

\section{Compiling ONNX Models}
\label{sec:onnx-mlir}
\begin{figure}[t]
   \centering
   \includegraphics[scale=0.5]{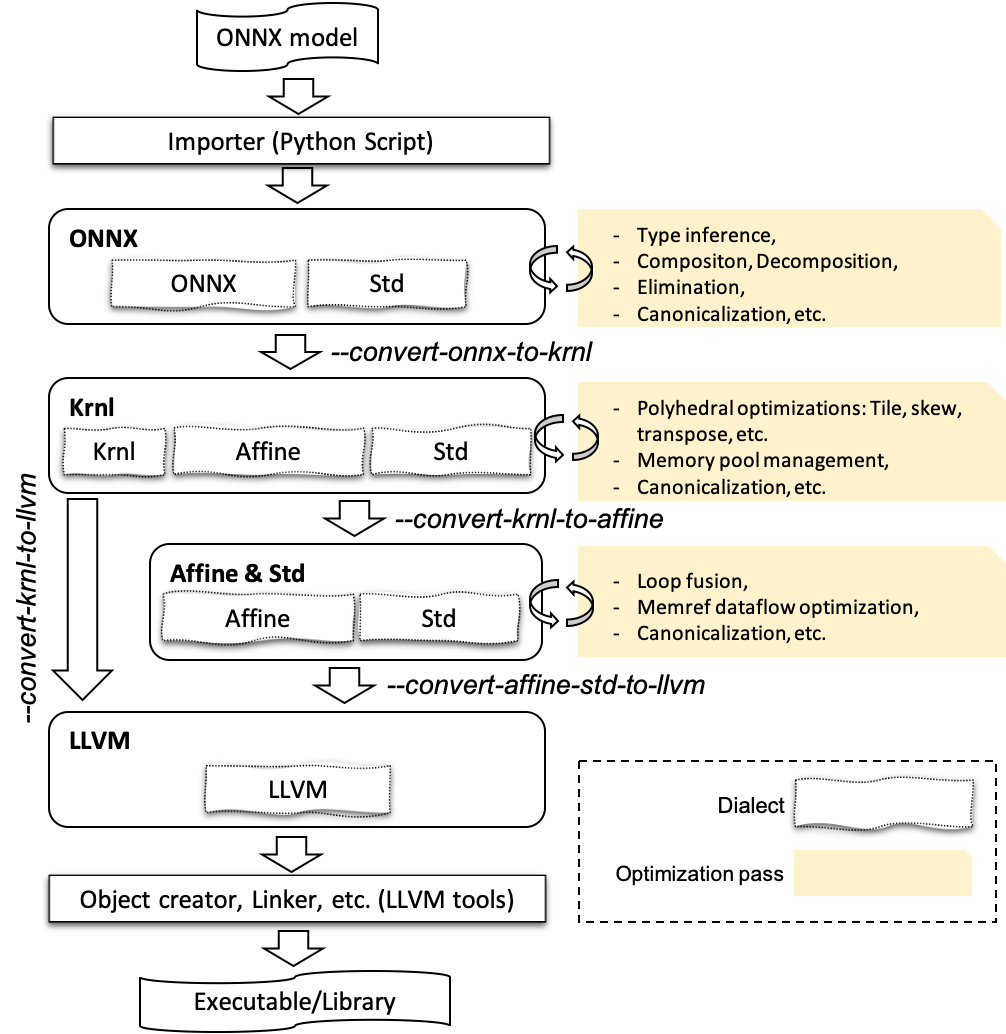}
   \caption{Architecture of onnx-mlir. Names prefixed with `\pass{}' are passes.}
   \label{fig:onnx-mlir}
\end{figure}

\begin{figure}[t]
   \centering
   \includegraphics[scale=0.4]{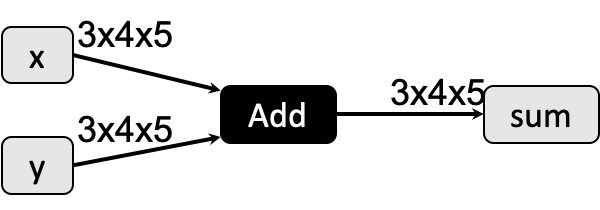}
   \caption{ONNX model for element-wise addition.}
   \label{fig:onnx-add}
\end{figure}

This section introduces our compiler, \emph{onnx-mlir}.
We first discuss its overall architecture.
We then introduce two new dialects, \dialect{onnx} and \dialect{krnl} dialects.
Finally, we present MLIR passes for carrying out optimization.

\subsection{Overview}
Figure~\ref{fig:onnx-mlir} shows the overall architecture of \onnxmlir.
The input is an ONNX model, and the output is a library containing the compiled code.
The output library contains an entry function called `\_dyn\_entry\_point\_main\_graph' whose inputs and outputs are similar to the ONNX model's inputs and outputs, respectively.
To carry out inference with the output library, users write their program to call the entry function by passing inputs to the function and obtain results.

There are five main dialects in \onnxmlir, i.e., \dialect{onnx}, \dialect{krnl}, \dialect{affine}, \dialect{std} and \dialect{llvm}, organized into four abstraction levels.
Two new dialects, \dialect{onnx} and \dialect{krnl}, are discussed in Sections ~\ref{sec:onnxir} and~\ref{sec:krnlir}, respectively.
The first abstraction level is a high-level representation of ONNX operations.
It consists of operations in \dialect{onnx} and \dialect{std} dialects, where the \dialect{onnx} dialect is automatically generated via \emph{an importer} that is a python script.
The second abstraction level includes \dialect{krnl}, \dialect{affine} and \dialect{std} dialects.
\dialect{krnl} dialect provides a representation that is suitable for loop optimizations, which is able to carry out affine transformations such as tile, skew, and permutation easily.
It plays as an intermediate dialect for efficiently lowering the \dialect{onnx} dialect into low-level dialects (e.g., \dialect{affine}, \dialect{std} and \dialect{llvm}).
The third abstraction level includes \dialect{affine} and \dialect{std} dialects where existing optimization passes in MLIR can be freely applied.
The forth abstraction level includes only \dialect{llvm} dialect that is realy to generate bitcode.

There are MLIR passes for converting one dialect to another, and for doing optimizations at a specific dialect.
\dialect{onnx} dialect is converted to \dialect{krnl} dialect via pass \pass{convert-onnx-to-krnl}.
Then \dialect{krnl} dialect (except some of its operations) is converted into \dialect{affine} and \dialect{std} dialects via pass \pass{convert-krnl-to-affine}.
The remaining operations in \dialect{krnl} dialect and operations in \dialect{affine} and \dialect{std} dialects are directly converted into instructions in \dialect{llvm} via pass \pass{convert-krnl-to-llvm}.
The right side of Fig.~\ref{fig:onnx-mlir} shows optimization passes that can be carried out at each abstraction level.

We only enumerate the important optimizations here, and the list of optimization passes is not exhaustive.

\begin{figure*}[!t]
  \begin{minipage}{\textwidth}
\begin{lstlisting}[style=mlir-code-inpara, caption={Operation \op{add} in \dialect{onnx} dialect, generated using importer.}, label={lst:add-onnxir}, frame=lines]
module {
  func @main_graph( %arg0:tensor<3x4x5xf32>, %arg1:tensor<3x4x5xf32>) -> tensor<*xf32> {
    %0 = "onnx.Add"(%arg0, %arg1) : (tensor<3x4x5xf32>, tensor<3x4x5xf32>) -> tensor<*xf32>
    std.return %0 : tensor<*xf32>
  }
  "onnx.EntryPoint"() {func = @main_graph, numInputs = 2 : i32, numOutputs = 1 : i32} : () -> ()
}
\end{lstlisting}

\begin{lstlisting}[style=mlir-code-inpara, caption={Operation \op{add} in \dialect{krnl} dialect, generated by applying passes \pass{shape-inference} and \pass{convert-onnx-to-krnl}.}, label={lst:add-krnlir}, frame=lines]
module {
  func @main_graph(%arg0: memref<3x4x5xf32>, %arg1: memref<3x4x5xf32>) -> memref<3x4x5xf32> {
    %0 = alloc() : memref<3x4x5xf32>
    %1:3 = krnl.define_loops 3
    krnl.iterate(%1#0, %1#1, %1#2) with (%1#0 -> %arg2 = 0 to 3, %1#1 -> %arg3 = 0 to 4, %1#2 -> %arg4 = 0 to 5) {
      %2 = affine.load %arg0[%arg2, %arg3, %arg4] : memref<3x4x5xf32>
      %3 = affine.load %arg1[%arg2, %arg3, %arg4] : memref<3x4x5xf32>
      %4 = std.addf %2, %3 : f32
      affine.store %4, %0[%arg2, %arg3, %arg4] : memref<3x4x5xf32>
    }
    std.return %0 : memref<3x4x5xf32>
  }
  "krnl.entry_point"() {func = @main_graph, numInputs = 2 : i32, numOutputs = 1 : i32} : () -> ()
}
\end{lstlisting}

\begin{lstlisting}[style=mlir-code-inpara, caption={Operation \op{add} in \dialect{affine} dialect, generated by applying the pass \pass{convert-krnl-to-affine}.}, label={lst:add-affinestdir}, frame=lines]
module {
  func @main_graph(%arg0: memref<3x4x5xf32>, %arg1: memref<3x4x5xf32>) -> memref<3x4x5xf32> {
    %0 = alloc() : memref<3x4x5xf32>
    affine.for %arg2 = 0 to 3 {
      affine.for %arg3 = 0 to 4 {
        affine.for %arg4 = 0 to 5 {
          %1 = affine.load %arg0[%arg2, %arg3, %arg4] : memref<3x4x5xf32>
          %2 = affine.load %arg1[%arg2, %arg3, %arg4] : memref<3x4x5xf32>
          %3 = std.addf %1, %2 : f32
          affine.store %3, %0[%arg2, %arg3, %arg4] : memref<3x4x5xf32>
        }
      }
    }
    std.return %0 : memref<3x4x5xf32>
  }
  "krnl.entry_point"() {func = @main_graph, numInputs = 2 : i32, numOutputs = 1 : i32} : () -> ()
}
\end{lstlisting}
  \end{minipage}
\end{figure*}
Before discussing dialects and optimization passes in detail, we give a brief running example and go through dialects in \onnxmlir.
This example is a testcase model in ONNX that performs element-wise binary addition.
Figure~\ref{fig:onnx-add} shows this ONNX model of the testcase.
Operation \op{add} accepts two tensors of type \shape{3x4x5xf32} (element type is float $32$) and returns a result tensor, i.e., sum, of the same type.
Listings~\ref{lst:add-onnxir},~\ref{lst:add-krnlir}, and~\ref{lst:add-affinestdir} show emitted programs in different dialects \dialect{onnx}, \dialect{krnl}, \dialect{affine}, respectively.
We omit the program in \dialect{llvm} due to space limitations.

In \dialect{onnx} dialect, operations are represented similarly to their descriptions in ONNX.
The ONNX model is converted into the function \op{main\_graph}.
To generate an entry point function into which users feed their inputs, we create a helper operation in the \dialect{onnx} dialect, i.e., \diop{onnx}{EntryPoint}, which keeps meta-data in the operation's attributes such as function name to call and the number of inputs and outputs. 

In \dialect{krnl} dialect, operation \diop{onnx}{Add} is translated into a loop-based computation represented by operations in the \dialect{krnl} dialect, where scalar computation is represented by primitive operations in the \dialect{affine} and \dialect{std} dialects. 
We can apply loop optimizations, such as tile, skew, or transpose, to loop-based computation.
At this level, we allocate memory for output tensors, and memory management can be performed.

In \dialect{affine} dialect, optimized loop-based computation in \dialect{krnl} dialect is translated into \diop{affine}{for} loops.
At this level, we still have an operation in \dialect{krnl}, i.e., \diop{krnl}{entry\_point}.
Such an operation is not related to the main computation and will be directly converted to \dialect{llvm}.
Operations in the \dialect{affine} dialect will be converted to operations in the \dialect{std} and \dialect{scf} dialects before being lowered to instructions in the \dialect{llvm} dialect.

\subsection{\dialect{onnx} dialect}
\label{sec:onnxir}

\begin{lstlisting}[style=mlir-code, caption={Tablegen-based definition for operation \op{relu}.}, label={lst:relu-def}, float=*t]
def ONNXLeakyReluOp:ONNX_Op<"LeakyRelu",
  [NoSideEffect, DeclareOpInterfaceMethods<ShapeInferenceOpInterface>]> {
    let summary = "ONNX LeakyRelu operation";
    let description = [{"LeakyRelu takes ... "}];
    let arguments = (ins AnyTypeOf<[TensorOf<[F16]>, TensorOf<[F32]>, TensorOf<[F64]>]>:$X, DefaultValuedAttr<F32Attr, "0.01">:$alpha);
    let results = (outs AnyTypeOf<[TensorOf<[F16]>, TensorOf<[F32]>, TenorOf<[F64]>]>:$Y);
    let extraClassDeclaration = [{ ... }];
}
\end{lstlisting}
\dialect{onnx} dialect is the first abstraction level in {\onnxmlir} and represents an ONNX model in MLIR language.
We wrote a python script to automatically import ONNX operations into the tablegen-based operation definitions in MLIR.
These imported operations are organized into the \dialect{onnx} dialect.
Thanks to tablegen, the operation definition in the \dialect{onnx} dialect is quite similar to the operation description in ONNX, where we are able to represent all necessary information, such as inputs, outputs, attributes, and description, into a single tablegen-based definition in human-readable textual form.

We also created a new operation in the \dialect{onnx} dialect, i.e., \diop{onnx}{EntryPoint} to keep information related to the dynamic list of inputs in an ONNX model.
This operation will be lowered to generate the entry function `\_dyn\_entry\_point\_main\_graph' of the generated library. 

Listing~\ref{lst:relu-def} shows a tablegen-based definition for the \op{relu} operation imported via the importer in \onnxmlir.
The operation description is represented in the `description' field (Line~$4$).
Inputs and attributes are represented in the `arguments' field, while outputs were represented in the `results' field (Lines~$5$--$6$).
All inputs and outputs will be imported as a tensor in MLIR.
The importer automatically infers element types for inputs, attributes, and outputs.
However, the shape of a tensor will be inferred via the \pass{shape-inference} pass, which is a trait in the \op{LeakyRelu} operation (Line~$2$).
MLIR generates a C++ class definition for an operation from its tablegen-based definition.
If users want to define custom declaration in the class, it can be done via the `extraClassDeclaration' field (Line~$7$).

\subsection{\dialect{krnl} dialect}
\label{sec:krnlir}

A computation kernel in a neural network workload has local structural simplicity in which loop nests are often simple, e.g., hyper-rectangle and statements carry quite straightforward arithmetic semantics. 
Such a characteristic is quite suitable to be represented in a polyhedral model for optimization~\cite{pouchet2008:iterative}.
\dialect{krnl} dialect aims to host both loop optimization and scalar semantic optimization in a single representation.
It is expected to provide interpretability where not only is polyhedral representation readable but it also makes program semantics (or what to execute) and program schedules (how and when to execute) independent.
In other words, our goal is to optimize not only programs but also the composition of individual schedules, which is a feature that is often lacking in other existing systems. 

Below is an example that defines a nested loop in \dialect{krnl}: 
\begin{lstlisting}[style=mlir-code-inpara]
%ii, %jj = krnl.define_loops 2
krnl.iterate(%ii, %jj) with (%ii -> %i = 0 to 10, %jj -> %j = 0 to 10) {
  %foo = std.addi %i, %j : index
}
\end{lstlisting}

{\noindent}where \diop{krnl}{define\_loops} defines \emph{two loops}, called {\sf ii} and {\sf jj}. 
These loop variables will be used to express both program semantics and schedules.
Operation \diop{krnl}{iterate} semantically accepts two types of loop variables: variables for original loops and variables for scheduled loops.
In syntactic sugar form, we separate the two types of loops by the keyword {\sf with}, i.e. (scheduled loops) {\sf with} (original loops).
Induction variables, e.g., {\sf i} and {\sf j} in the above example, will be defined by using original loops.
If there is no schedule (e.g. block, skew, etc.), the scheduled loops are similar to the original loops.

Now, we insert a schedule for blocking or tiling.
Without loss of generality, we define just one loop instead of two.
\begin{lstlisting}[style=mlir-code-inpara]
%ii = krnl.define_loops 1
%ib, %il = krnl.block %ii 2 : (!krnl.loop)->(!krnl.loop, !krnl.loop)
krnl.iterate(%ib, %il) with (%ii -> %i = 0 to 10) {
  %foo = std.addi %i, %i : index
}
\end{lstlisting}

Operation \diop{krnl}{block} (Line~$2$) takes a loop and integer as inputs, where the integer is the tile size with which we want to carry out blocking.
Results are two loop variables: one for the outer loop and the other for the inner loop.
The two loops will be used as the result of scheduling and be passed to \diop{krnl}{iterate} (Line~$3$).
It is worth noting that the original loops and computation in \diop{krnl}{iterate} remained \emph{unchanged} while inserting a schedule, which is exactly what we want for seperating program semantics and schedules in our \dialect{krnl} dialect.

The \pass{convert-krnl-to-affine} pass automatically generates optimized \diop{affine}{for} based loops as follows.
\begin{lstlisting}[style=mlir-code-inpara]
#map0 = affine_map<(d0) -> (d0)>
#map1 = affine_map<(d0) -> (d0 + 2)>
affine.for %arg0 = 0 to 10 step 2 {
  affine.for %arg1 = #map0(%arg0) to #map1(%arg0) {
    %0 = addi %arg1, %arg1 : index
  }
}
\end{lstlisting}

{\noindent}The outer \diop{affine}{for} iterates with step $2$ i.e., the tile size, and the inner \diop{affine}{for} iterates over the elements in a tile.

Other schedules, such as skew and permutation are used in a similar manner.
All schedules are composable and can be nested.

\subsection{Optimization Passes}
\label{sec:opt-pass}
In this section, we discuss some of the optimization passes in \onnxmlir.
Thanks to the expressive power of MLIR, many optimizations can be expressed easily via Declarative Rewriting Rules (DRRs) using tablegen records or writing code in C++.

\subsubsection{Operation Decomposition}
In ONNX, many operations can be expressed using other basic operations.
For example, \op{ReduceL1} over a vector $x$ is mathematically calculated by summing up the absolute values of the elements in $x$.
In other words, we have
$$
\op{ReduceL1} = \op{ReduceSum}\ (\op{Abs}\  x)
$$

We only need to lower a subset of operations in the \dialect{onnx} dialect to \dialect{krnl} dialect, while the remaining operations in the \dialect{onnx} dialect will be decomposed into operations in the subset. 

Using the DRRs in MLIR, operation decomposition is concisely written as the following pattern: 
\begin{lstlisting}[style=mlir-code-inpara]
def ReduceL1Pattern: Pat<
  (ReduceL1Op $x, $axes, $keepdims),
  (ReduceSumOp (AbsOp $x), $axes, $keepdims)
>;
\end{lstlisting}

{\noindent}where \op{ReduceL1Op}, \op{ReduceSumOp}, and \op{AbsOp} are programmable forms of operations \diop{onnx}{ReduceL1}, \diop{onnx}{ReduceSum}, and \diop{onnx}{Abs} respectively.
Variables $\sf{x}$, $\sf{axes}$, and $\sf{keepdims}$ are for keeping input values of operation \op{ReduceL1Op}.
The pattern `ReduceL1Pattern' contains a source pattern to match a graph of one operation \op{ReduceL1Op} (Line~$2$) and a destination pattern to generate a graph of two operations \op{ReduceSumOp} and \op{AbsOp} (Line~$3$).
Whenever an operation \op{ReduceL1Op} appears in an ONNX model, it will be replaced with a combination of \op{ReduceSumOp} and \op{AbsOp}. 

\subsubsection{Shape Inference}
The \pass{shape-inference} pass attempts to infer shapes for all tensors in a program at \dialect{onnx}.
The pass traverses all operations in a program, infers the shapes of tensors with unrank shapes (i.e. \tensor{*xf32}), propagates the ranked shapes to consuming operations, and terminates once all tensors have ranked shapes.
For one operation, if its inputs have static shapes, it is likely that the \pass{shape-inference} pass will be able to infer static shapes for its outputs.
If the inputs have dynamic shapes (e.g. \tensor{?x?x?xf32}), the outputs will also have dynamic shapes also, except for some operations whose output tensors' shapes are specified in the operation attributes.

\subsubsection{Graph Rewriting}
\label{sec:graph-rewriting}
Graph rewriting is a powerful optimization tool.
It is intensively applied to neural networks since calculation in a neural network is expressed via a dataflow graph. 
In MLIR, graph rewriting rules are conveniently represented using DRRs.

For example, the following rule is to fuse \diop{onnx}{Add} and \diop{onnx}{MatMul} into a single operation \diop{onnx}{Gemm} under the condition that the result of \op{MatMulOp} is only consumed by \op{AddOp}: 

\begin{lstlisting}[style=mlir-code-inpara]
def MulAddToGemmPattern : Pat<
  (AddOp (MatMulOp:$res $m1, $m2), $m3), 
  (GemmOp $m1, $m2, $m3),
  [(HasOneUse $res)]
>;
\end{lstlisting}

Another example is to remove an \op{IdentityOp} operation by passing its input directly to its consuming operations.
\begin{lstlisting}[style=mlir-code-inpara]
def IdentityEliminationPattern : Pat<
  (ONNXIdentityOp $arg),
  (replaceWithValue $arg)
>;
\end{lstlisting}

Users can write as many rewriting rules as possible in the same manner.

\subsubsection{Constant propagation}
Constant propagation is a well-known optimization in compilers.
In \onnxmlir, we created a pass to do this during compilation.
There are two key ideas in constant propagation: 
\begin{inparaenum}
  \item if all the inputs of an operation are constant, compute its outputs during compilation and remove the operation,
  \item if there is a mix of constant and non-constant inputs, normalize the operation.
\end{inparaenum}
Normalization is to increase the possibility of constant propagation and strongly depends on the mathematical properties of an operation.
Below are some normalization rules in {\onnxmlir} for the \diop{onnx}{Add} operation whose properties are associative and communicative.
\begin{enumerate}
  \item $c + x \Rightarrow x + c $
  \item $(x + c_1) + c_2 \Rightarrow x + (c_1 + c_2) $
  \item $(x + c) + y \Rightarrow (x + y) + c $
  \item $x + (y + c) \Rightarrow (x + y) + c $
  \item $(x + c_1) + (y + c_2) \Rightarrow (x + y) + (c_1 + c_2) $
\end{enumerate}
where $x$ and $y$ are non-constant values, and $c$, $c_1$, and $c_2$ are constant values.
Normalization rules are expressed by using the DRRs in MLIR.


\section{Preliminary Experiments}
\label{sec:experiment}
\subsection{ONNX operation support and testcases}
ONNX provides a set of test cases for each operation.
When we support any operation in {\onnxmlir}, we enable its ONNX test cases to check whether the operation behaves correctly and produces correct result.
At the time of writing this paper, {\onnxmlir} supports $51$ operations out of $139$ operations in ONNX, including important operations such as convolution, pooling, Gemm, and LSTM.
These are enough to compile and execute major networks such as MNIST and ResNet50.
On the GitHub repository of {\onnxmlir}, we enable continuous integration on different environments, i.e., Windows, Linux, and Docker environments, and different systems, i.e., x86 machines, IBM Power Systems, and System Z.
All supported operations have passed tests on the above environments.

\subsection{MNIST and ResNet50}
In this section, we present some of our preliminary results for two neural network models in the ONNX Model Zoo: MNIST and ResNet50~\cite{He:2015:CORR}.
The MNIST\footnote{\url{https://github.com/onnx/models/tree/master/vision/classification/mnist}} and ResNet50\footnote{\url{https://github.com/onnx/models/tree/master/vision/classification/resnet}} models have already been trained in the CNTK and Caffe2 frameworks, respectively.
We ran inferences on the given test data set in each model.
The experiments were conducted on a machine with 2.3-GHz POWER9 processors.
For {\onnxmlir}, graph rewriting and canonicalization passes were enabled.
%
%
%
In this paper, we only provide a reference implementation that is not optimized, thus performance measurements are not applicable.

\begin{table}[!h]
  \centering
  \caption{Run inferencing with MNIST and ResNet50 on a POWER9 machine. Time in seconds.}
  \begin{tabular}{ccc}
    \toprule
    Model & Compilation time & Inference time\\
    \midrule
    MNIST & 0.237 & 0.001 \\
    ResNet50 & 7.661 & 7.540 \\
    \bottomrule
  \end{tabular}
  \label{tab:inference-mnist-restnet50}
\end{table}

Table~\ref{tab:inference-mnist-restnet50} shows the running times for the MNIST and ResNet50 models when doing inferencing.
For each model, we measured the compilation time for compiling the model to native code and inference time for running the native code with real inputs.
MNIST is a small model with two convolutional operations, one max pooling operation and a matrix multiplication followed by an element-wise addition.
Compiling the MNIST model and carrying out inferencing was rather fast, i.e., finished in less than one second.
In the MNIST model, the graph rewriting rule {\sf MulAddToGemmPattern} mentioned in Sec.~\ref{sec:graph-rewriting} was applied to fuse matrix multiplication and element-wise addition into a Gemm operation.
ResNet50 is a complex deep model consisting of $50$ layers of operations such as convolutions and poolings.
The model is about $100$ megabytes including learned weights.
For ResNet50, the current version of {\onnxmlir} does not have any optimization applied to the model during compilation.
However, we believe that the compilation time looks reasonable and the inference time is not so slow.
We hope that once we integrate important optimizations, such as polyhedral optimizations, SIMD optimization, and loop fusion in near future, the inference time will be significantly reduced.

\subsection{Supported Systems}
Although {\onnxmlir} is completely built upon widely-used open source software such as ONNX and MLIR,
we found a problem related to supporting different systems.
In particular, we could not run ONNX models on Linux on IBM System Z (s390-linux) because the big-endian format was not well-supported in ONNX and MLIR.
There are two reasons for such a problem.
First, a large amount of public input data and models in ONNX are stored in little-endian format.
Hence, they must be converted to big-endian format before they are used in a big-endian system.
Second, we found that constant values in ONNX models are not correctly loaded in MLIR.
LLVM was well-supported in big-endian, but MLIR was not.
We created two patches to solve this problem: one in ONNX\footnote{\url{https://github.com/onnx/onnx/pull/2633}} and one in MLIR\footnote{\url{https://reviews.llvm.org/D78076}}, and they are now available at the master branches of ONNX and MLIR. 
As a result, {\onnxmlir} now supports Linux on x86 (x86-Linux), Linux on Power Systems (ppc64le-Linux), Linux on IBM Z (s390-Linux), and Windows.

\section{Conclusion}
\label{sec:conclusion}
We are developing an open source compiler called {\onnxmlir} for compiling ONNX models into native code.
MLIR was used as an infrastructure to build the compiler, and two novel dialects were introduced, i.e., \dialect{onnx} and \dialect{krnl}.
We also discussed some optimizations such as graph rewriting and constant propagation.
It is worth noting that new optimizations can be easily integrated into {\onnxmlir} thanks to the MLIR infrastructure. 
In the future, we will add more optimizations, e.g., polyhedral optimization, loop fusion, SIMD optimization, and enable code generation for accelerators.

\section*{Acknowledgements}
We acknowledge the ONNX standard for hosting the project and the external contributors for their contributions.

\balance  
\bibliographystyle{ipsjsort-e}
\bibliography{refs}

\end{document}